\documentclass[aps,prb,superscriptaddress,a4paper,11pt]{revtex4-1}

\usepackage[english]{babel}
\usepackage{graphicx}
\usepackage{hyperref}
\usepackage{amsmath}
\usepackage{amssymb}
\usepackage{color,soul}

\DeclareMathOperator{\ce}{ce}
\DeclareMathOperator{\se}{se}
\DeclareMathOperator{\me}{me}

\DeclareMathOperator{\Tr}{Tr}

\DeclareMathOperator{\Arg}{Arg}

\newcommand{\buenosaires}{Departamento de F\'isica and IFIBA, FCEN, Universidad de Buenos Aires, 
Ciudad Universitaria, Pab.\ I, C1428EHA Buenos Aires, Argentina}

\begin{document}

\title{Analytical solution of narrow quantum rings with general Rashba and Dresselhaus spin-orbit couplings}

\author{J.\ M.\ Lia}
\affiliation{\buenosaires}

\author{P.\ I.\ Tamborenea}
\affiliation{\buenosaires}



\date{\today}

\begin{abstract}
We solve analytically the energy eigenvalue problem of narrow semiconductor quantum rings
with a general spin-orbit term that includes as a special case the Rashba and Dresselhaus
interactions acting simultaneously.
The eigenstates and eigenenergies of the system are found for arbitrary values of the
spin-orbit coupling constants without making use of approximations.
The general eigenstates are expressed as products of a scalar Mathieu function and a
spinor factor which is periodic or pseudo-periodic on the ring.
Our general solution reduces to the previously found solutions for particular combinations
of the Rashba and Dresselhaus couplings, like the well-studied cases of Rashba-only and of
equal coupling constants.
%
\end{abstract}

\maketitle

\section{Introduction}


Semiconductor quantum rings (QR) are an elegant example of how fundamental quantum mechanical 
systems can be realized with current nanofabrication techniques.\cite{fom}
Their simple quasi-one-dimensional circular geometry lends itself perfectly to the study 
of orbital angular momentum of charge carriers.
In addition, the spin degree of freedom can be brought into play thanks to the Rashba (RSOI) 
and Dresselhaus (DSOI) spin-orbit interactions, present and controllable in semiconductor nanostructures.
Furthermore, the injection of angular momentum via excitation with twisted light can provide 
a way to initialize the system in a finite-angular-momentum state.\cite{cyg-tam-axt,qui-tam-ber}
Thus, the stage is set for an interesting angular momentum dynamics to take place, which
could be studied and ultimately controlled for its use in quantum information processing.
From a fundamental point of view, QR are an excellent scheme for studying spin-interference
quantum mechanical effects.\cite{koe-tsc-han,ber-kob-sek,nag-fum-fru}

It has been recognized that the simultaneous action of RSOI and DSOI in semiconductor 
nanostructures can lead to interesting and varied effects.\cite{sch-egu-los, rom-ull-tam, mai}
Recently, several authors have investigated this combination of spin-orbit interactions for 2D 
and 1D quantum rings.\cite{sha-sza-esm,zam-aza-nik,pou-rez}
Up to now, the necessary electronic structure---eigenvalues and eigenstates of the one-electron
Hamiltonian in the presence of both RSOI and DSOI, and eventually also of an external magnetic field---used
in transport and optical-properties studies has usually been obtained numerically.
The only exception in which the analytic solution has also been found is the case with equal 
RSOI and DSOI coupling constants.\cite{sch-egu-los}
On the other hand, the problem with RSOI alone in 1D QR has been solved analytically by Frustaglia 
and Richter,\cite{fru-ric} while, to the best of our knowledge, no analytical solution 
to the problem with DSOI alone has been given in the literature.
In this article we provide an analytical solution to the eigenvalue problem of the Hamiltonian 
of a quasi-1D QR with both RSOI and DSOI of arbitrary intensity.

Actually, the Hamiltonian of the DSOI in quantum wells is not unique since it depends 
on the orientation of the quantum well relative to the crystal axes.\cite{win}
For this reason, we consider a generalized (linear-in-$k$) form for the spin-orbit coupling 
in quantum wells.
The solution provided here applies to this general problem.

This article is organized as follows.
In Section \ref{sec:system} we describe the quantum-ring system with the general spin-orbit
interaction which includes the Rashba and Dresselhasus interactions.
In Section \ref{sec:solution} we provide the derivation of the analytical solution to
the general spin-orbit problem.
In Section \ref{sec:free-electron} we briefly review the solution of the ring without spin-orbit 
interaction and in Sections \ref{sec:only-alpha} and \ref{sec:equal} we recover the known solutions 
of the ring with RSOI alone and with RSOI and DSOI with equal coupling constants, respectively.
In Section \ref{sec:general-case} we explore the solution with general RSOI and DSOI acting
simultaneously, and in Section \ref{sec:conclusion} we provide concluding remarks.


\section{Ring System}
\label{sec:system}

We consider a narrow homogeneous semiconductor QR of inner radius $a$. 
We assume that the QR is doped with a conduction-band electron which is subject to spin-orbit interaction. 
We work in the envelope-function approximation and with a Hamiltonian of the form
%
%
%
\begin{equation}
  H = H_0 + H_R + H_D + H_{\Delta},
  \label{eq:Hamiltonian_Generic}
\end{equation}
where $H_0$ is the envelope-function Hamiltonian without the spin-orbit interactions,
$H_R$ and $H_D$ are respectively the well-known Rashba and Dresselhaus spin-orbit
Hamiltonians
\begin{align}
  H_R &= \alpha(k_x\sigma_y - k_y\sigma_x), \\
  H_D &= \beta(k_x\sigma_x - k_y\sigma_y),
\end{align}
and the last term
\begin{equation}
  H_{\Delta} = (\delta_{xx}k_y + \delta_{yx}k_y)\sigma_x
\end{equation}
represents deviations from these two interactions. 
In the expressions above, $k_x=-i\partial_x$ and $k_y = -i\partial_y$ 
are momentum operators in coordinate space, $\sigma_x$ and $\sigma_y$ are the Pauli 
matrices and $\alpha,\beta,\delta_{xx}$ and $\delta_{yx}$ are real constants that 
depend only on the properties of the system.



We set the QR on the $xy$-plane centered at the origin of coordinates and write
\begin{equation}
  H_0 = -\hbar^2\nabla^2/2m^\ast + V(\mathbf{r}),
\end{equation}
with $m^\ast$ the conduction-band effective mass and $V(\mathbf{r})$ the confining potential
that defines the QR.
%
%
Switching to cylindrical coordinates and introducing the operators
\begin{equation}
  \partial^\pm = \partial_x\pm i\partial_y = e^{\pm
    i\phi}\left(\partial_r\pm\frac{i}{r}\partial_\phi\right),
\end{equation}
the Hamiltonians $H_R$, $H_D$ and $H_{\Delta}$ can be recast into the forms
\begin{align}
  H_R &= \alpha(\partial^-S^+ - \partial^+S^-), \\
  H_D &= -i\beta(\partial^-S^- + \partial^+S^+), \\
  H_{\Delta} &= (\Delta^\ast\partial^- - \Delta\partial^+)(S^+ + S^-)
\end{align}
where $\Delta = (\delta_{yx} + i\delta_{xx})/2$ and \(S^\pm = \sigma_x \pm i\sigma_y\) are
the spin raising (+) and lowering ($-$) operators.

We now take the limit of very narrow, quasi-one-dimensional QR and, following
Ref.\ [\onlinecite{mei-mor-klap}], assume that the radial-part contributions to the
eigenfunctions of $H$, arising from the finite width of the ring, are adequately described
by the lowest radial eigenfunction $R_0(r)$ of $H_0$.
In line with this assumption, we replace factors $1/r$ and terms involving
first-derivatives in $r$, appearing in both the azimuthal part of $H_0$ and the operators
$\partial^{\pm}$, with the quantities $\langle 1/r\rangle_{R_0}\approx 1/a$ and
$\langle\partial_r\rangle_{R_0} = \int R_0(r)\partial_rR_0(r) rdr \approx -1/2a$,
respectively.

%
We thus arrive at the following effective one-dimensional ($\phi$-dependent) Hamiltonian:
\begin{equation}
  \begin{aligned}
    H_{\phi} = &-\epsilon_0\partial_{\phi}^2 + \left[(\alpha - \Delta)S^- - (\Delta +
      i\beta)S^+\right]\frac{e^{i\phi}}{a}\left(i\partial_{\phi} - \frac{1}{2}\right) + \\ 
    &\left[(\alpha - \Delta^\ast)S^+ - (\Delta^\ast -
      i\beta)S^-\right]\frac{e^{-i\phi}}{a}\left(i\partial_{\phi} + \frac{1}{2}\right);
  \end{aligned}
  \label{eq:azi-hamil}
\end{equation}
where $\epsilon_0 = \hbar^2/2m^\ast a^2$.

We define the $2\pi$-periodic, anti-hermitian operator:
\begin{equation}
  F(\phi) = \frac{i}{a}\left\{e^{i\phi}\left[(\alpha - \Delta)S^- - (\Delta +
      i\beta)S^+\right] + e^{-i\phi}\left[(\alpha - \Delta^\ast)S^+ - (\Delta^\ast -
      i\beta)S^-\right]\right\}
  \label{eq:F-def}
\end{equation}
in terms of which the Hamiltonian reads
\begin{equation}
  H_{\phi} = -\epsilon_0\partial_{\phi}^2 + F\partial_{\phi} +
  \frac{1}{2}(\partial_{\phi}F)
\end{equation}


\section{General analytic solution}
\label{sec:solution}

Our goal is to solve the spinor eigenvalue problem
\begin{equation}
  H_{\phi}\eta(\phi) = E\eta(\phi).
  \label{eq:eigen-prob}
\end{equation}
We propose a factorised solution of the form $\eta(\phi) = f(\phi)\chi(\phi)$, where $f$
is a complex-valued scalar function and $\chi(\phi)$ a complex-valued spinor, both to be
determined.
Inserting the proposed solution into Eq. \eqref{eq:eigen-prob}, we get
\begin{equation}
  \epsilon_0(f''\chi + f\chi'') + (2\epsilon_0\chi' - F\chi)f' - Ff\chi' +
  \left(-\frac{F'}{2} + E\right)f\chi = 0 ,
  \label{eq:f-chi}
\end{equation}
where the primes denote derivatives with respect to $\phi$.
As we have imposed no restriction on the form or properties of the factors $f(\phi)$ and
$\chi(\phi)$, aside from the basic requirement of being smooth functions of $\phi$, 
we can conveniently pick $\chi$ from among the solutions to the equation
\begin{equation}
  2\epsilon_0 \chi' - F(\phi)\chi = 0 .
  \label{eq:chi}
\end{equation}
%

Assuming that $\chi(\phi)$ is not identically zero, this choice reduces
Eq. \eqref{eq:f-chi} to a differential equation for $f(\phi)$ alone:
\begin{equation}
  \epsilon_0f'' +
  \left(-\frac{|\Gamma|}{2a^2\epsilon_0}\cos(2\phi + \phi_{\Gamma}) +
  \frac{|\alpha - \Delta|^2 + |\Delta + i\beta|^2}{4a^2\epsilon_0} + E\right)f = 0
  \label{eq:f}
\end{equation}
%
%
where $\Gamma = (\alpha - \Delta)(\Delta + i\beta)$ and $\phi_{\Gamma} = \Arg\Gamma$.
%
%

It is important to bear in mind that the separation of Eq. \eqref{eq:eigen-prob} into 
an equation for each of the factors in $\eta(\phi) = f(\phi)\chi(\phi)$ does not indicate
that any product of solutions to Eqs. \eqref{eq:chi} and \eqref{eq:f} form an eigenstate
of $H_{\phi}$.
Indeed, both factors $f(\phi)$ and $\chi(\phi)$ are still related through the energy
eigenvalue $E$, which enters Eq. \eqref{eq:f} as a parameter with no restriction other
than being real, and is ultimately determined by imposing the condition that its
associated eigenstate $\eta(\phi)$ be single-valued on $\phi$.


Let us study each of the Eqs. \eqref{eq:chi} and \eqref{eq:f} separately. It can be shown
(see Appendix A) that, irrespective of the quantities $\alpha,\beta,\Delta$ and
$\epsilon_0$, the former always has a set of two pointwise orthonormal solutions
$\chi_{\mu s}(\phi)$ which satisfy the pseudo-periodic property
\begin{equation}
  \chi_{\mu s}(\phi + 2\pi) = e^{is2\pi\mu}\chi_{\mu s}(\phi),
\end{equation}
where $s=\pm1$, $0\leq\mu\leq1/2$ and $e^{is2\pi\mu}$ is a characteristic Floquet
multiplier of Eq. \eqref{eq:chi}.
The latter, in turn, can be recast into the general form of the well-known and extensively
studied Mathieu equation\cite{mcl}
\begin{equation}
  (p - 2q\cos(2\phi))f + f'' = 0,
  \label{eq:mathieu}
\end{equation}
by applying the translation $2\phi + \phi_{\Gamma}\rightarrow 2\phi$ and defining the
dimensionless parameters
\begin{equation}
  \begin{aligned}
    2q &= \frac{|\Gamma|}{2a^2\epsilon_0^2} = \frac{|(\alpha-\Delta)(\Delta +
      i\beta)|}{2a^2\epsilon_0^2}, \\
    p &= \frac{|\alpha - \Delta|^2 + |\Delta +
      i\beta|^2}{4a^2\epsilon_0^2} + \frac{E}{\epsilon_0}.
  \end{aligned}
\end{equation}
It can be seen, on the one hand, that parameter $q$ is always real and that it is
completely determined by the quantities $\epsilon_0,\alpha,\beta$ and $\Delta$ that define
$H_{\phi}$.
On the other hand, the parameter $p$, unlike $q$, depends on the energy and therefore
it can be chosen freely, provided that it remains a real quantity.
%
%
This property is important since, with recourse to Mathieu's equation theory, it can be
shown \cite{mcl,wol} that if $q\in\mathbb{R}$ and $\nu$ is chosen real,
then there exist a real $p(\nu;q)$ and a solution $f_{\nu}(\phi;q)$ to
Eq. \eqref{eq:mathieu} associated with it that satisfies the pseudo-periodic property
\begin{equation}
  f_{\nu}(\phi + 2\pi;q) = e^{2\pi i\nu}f_{\nu}(\phi;q).
\end{equation}


This freedom in choosing $\nu$ suggests that any single-valued (i.e., periodic) eigenstate
of $H_{\phi}$ may be assembled from a
pseudo-periodic spinor $\chi_{\mu s}(\phi)$ and a Mathieu function $f_{\nu}(\phi; q)$ by
conveniently choosing the latter so that their product $f_{\nu}(\phi +
\phi_{\Gamma}/2;q)\chi_{\mu s}(\phi)$ satisfies
\begin{equation}
  \begin{aligned}
    f_{\nu}\left(\phi + \frac{\phi_{\Gamma}}{2} + 2\pi;q\right)\chi_{\mu s}(\phi + 2\pi) &=
    \left[e^{2\pi i\nu}f_{\nu}\left(\phi +
      \frac{\phi_{\Gamma}}{2};q\right)\right]\left[e^{2\pi i s\mu}\chi_{\mu
        s}(\phi)\right] \\
    &= f_{\nu}\left(\phi + \frac{\phi_{\Gamma}}{2};q\right)\chi_{\mu s}(\phi).
  \end{aligned}
  \label{eq:method}
\end{equation}
This requirement can be met by picking $\nu = -s(\mu - m)$, with $m\in\mathbb{Z}$.
The integer $m$ takes into account the fact that only the fractional part $-s\mu$ of the
Floquet exponent $\nu$ is unique, since adding an integer to it leaves its corresponding
Floquet multiplier invariant.
%

%

The relation between $\nu$ and $\mu$ thus defines a set of Mathieu functions
$f_{-s\mu+sm}(\phi;q)$ which can be shown to be orthonormal (see Appendix B) on
$0\leq\phi\leq2\pi$.
It also determines the energy spectrum $E/\epsilon_0$ in terms of $\epsilon_0$ and the SO
coupling constants through the associated set of values for the parameter $p$, $p(-s\mu +
sm;q)$.

In order to give a concrete expression $\eta_{\mu s,m}(\phi)$, we separate the pure
periodic case where $\mu = 0$ for which the solutions to Mathieu's equation can be chosen
to have well-defined parity with respect to $\phi$, from the pseudo-periodic ones where
$\mu\neq0$.
We thus write, for the former
\begin{equation}
  \eta_{\pm,m}(\phi;q) = f_m\left(\phi + \frac{\phi_{\Gamma}}{2}\right)\chi_{\pm}(\phi),
  \label{eq:eigenspinor-periodic}
\end{equation}
where $\chi_{\pm}(\phi)$ are two orthonormal solutions to Eq. \eqref{eq:chi} and
\begin{equation}
  f_m\left(\phi + \frac{\phi_{\Gamma}}{2};q\right) = \frac{1}{\sqrt{\pi}}
  \begin{cases}
    \ce_m\left(\phi + \frac{\phi_{\Gamma}}{2};q\right) & m\geq 0 \\
    -i\se_{-m}\left(\phi + \frac{\phi_{\Gamma}}{2};q\right) & m< 0,
  \end{cases}
  \label{eq:fm-periodic}
\end{equation}
with $\ce_m(\phi;q)$ and $\se_{-m}(\phi;q)$ the even and odd Mathieu functions of integer
order, respectively.
The expressions for the latter cases are, in turn,
\begin{equation}
  \eta_{\mu\pm,m}(\phi) = \frac{1}{\sqrt{2\pi}}\me_{\mp(\mu - m)}\left(\phi +
  \frac{\phi_{\Gamma}}{2};q\right)\chi_{\mu\pm}(\phi).
\end{equation}
In both the periodic and pseudo-periodic cases, the normalization of each eigenstate
depends only on the scalar factor $f_{-s\mu+sm}(\phi;q)$, as it can be seen by computing
the product $\eta_{\mu s,m}(\phi)^\dag\eta_{\mu s,m}(\phi)$ and recalling that the
spinors $\chi_{\mu s}(\phi)$ are pointwise orthonormal.


%
Closed analytical expressions for the spinors $\chi_{\pm\mu}(\phi)$ and the energy spectrum
$E/\epsilon_0$ can only be obtained for a handful of special cases, some of which have already
been completely\cite{fru-ric} o partially\cite{sch-egu-los} solved.
Nevertheless, expansions of $E/\epsilon_0$ in powers of $q$ are known
\cite{mcl,ars,tam-wan} for both the periodic and pseudo-periodic
cases.
For brevity, we only reproduce here the first few terms given in
Ref.\ [\onlinecite{wol}] for the latter case that approximate the spectrum when $|q|$
is small compared to unity
\begin{equation}
  E_m(\mu,q) = -\frac{\alpha^2 + \beta^2}{4\epsilon_0a^2} + \epsilon_0(\mu - m)^2 +
  \frac{\epsilon_0q^2}{2(\mu - m)^2 - 2} + O(q^4).
  \label{eq:spectrum}
\end{equation}
It is worth noticing that although neither of these terms have singularties when
$\mu\neq0$, this series may not be suitable for numerical estimations\cite{mcl}
of the spectrum when $\mu\rightarrow0$, as it converges slowly,\cite{tam-wan} even for
small $|q|$.
In those cases, other and more accurate methods are available.\cite{tam-wan,mcl,wol,sch-egu-los}
It stems from the expansion in Eq. \eqref{eq:spectrum} that, at least to fourth order in
$|q|$, the spectrum does not depend on the sign of the quantity $\mu - m$. This remarkable
property actually holds\cite{wol} to any order in $|q|$ and implies that the
eigenstates $\eta_{\mu\pm,m}(\phi;q)$ of $H_{\phi}$ in the cases for which $0<\mu$ are
degenerate.
Similarly, in the case $\mu=0$ the eigenstates $\eta_{\pm,m}(\phi;q)$ share the same
energy, since the spectrum is fundamentally independent of the spinor $\chi_{\pm}(\phi)$.
%
In every case, this degeneracy is the two-fold Kramers' degeneracy that arises from the
time-reversal symmetry of the total Hamiltonian $H$.\cite{sak-nap} Furthermore, it can be
shown by direct computation that, up to a constant phase, the eigenstates
$\eta_{\mu+,m}(\phi;q)$ and $\eta_{\mu-,m}(\phi;q)$ are one the time-reversed state of the
other.

In the next Section we revisit some of the special cases in which analytical
expressions for either the eigenstate $\eta_{\mu\pm,m}(\phi;q)$ or the spinor part
$\chi_{\mu s}(\phi)$ can be found.
It is our purpose to show that the general method developed above effectively reproduces
these well-known results.


\section{Special cases}

\subsection{Electron in the conduction band without SO}
\label{sec:free-electron}
In this case, the SO interaction is completely absent. The Hamiltonian that describes the
electron in the conduction band therefore reduces to
\begin{equation}
  H_0 = \frac{\epsilon_0}{\hbar^2}L_z^2 = -\epsilon_0\partial_{\phi}^2;
\end{equation}
which can be obtained by setting $\alpha=\beta=\Delta=0$ in $H_{\phi}$. 
Eigenfunctions of
$H_0$ that are single-valued on the ring can be readily obtained from the rightmost
equality and are of the form $e^{im\phi}\chi$, where $m\in\mathbb{Z}$ and $\chi$ is a
constant spinor, as $H_0$ is independent of spin. 
The scalar factors $e^{im\phi}$ in
this case are also periodic solutions to Mathieu's equation \eqref{eq:f}.

It can also be seen that the operator $F(\phi)$ vanishes in this case and that
Eq. \eqref{eq:chi} is reduced to $\chi'=0$. 
Solutions to this equation are constant
spinors which can be thought of as $2\pi$-periodic functions of $\phi$ and therefore
correspond to the case for which $\mu=0$. 
This result is thus consistent with the periodicity of the eigenstates of $H_0$.

The spectrum of $H_0$ can be readily obtained and takes the form $\epsilon_0m^2$, which
coincides with the zeroth-order term in $q$ in the expansion \eqref{eq:spectrum}.

\subsection{Rashba only}
\label{sec:only-alpha}

In this case, only the Rashba interaction is considered and the corresponding Hamiltonian
is obtained by setting $\beta = \Delta = 0$ in $H_{\phi}$.
In a previous work,\cite{fru-ric} it has been shown the eigenvalue problem
\eqref{eq:eigen-prob} is exactly solvable and that the eigenvectors have, in the basis of
eigenstates of $\sigma_z$, the general structure,
\begin{equation}
  \eta_m(\phi) = e^{im\phi}
  \begin{pmatrix}
    c_{\uparrow}\\c_{\downarrow}e^{i\phi}
  \end{pmatrix},
  \label{eq:rashba-exact}
\end{equation}
with $m\in\mathbb{Z}$ and $c_{\uparrow,\downarrow}\in\mathbb{C}$ constants dependent on
the parameters of the system. 
This eigenstate can be decomposed into a product of a scalar
function and a spinor, both pseudo-periodic in $\phi$, by rewriting it as follows
\begin{equation}
  \eta_{m}(\phi) = e^{i(m-\mu)\phi}\left[e^{i\mu\phi}\begin{pmatrix}
    \chi_{\uparrow}\\\chi_{\downarrow}e^{i\phi}
  \end{pmatrix}\right],
\end{equation}
with $\mu\in\mathbb{R}$ to be determined.
Since the parameter $q$ vanishes in this case, the factor $e^{i(m-\mu)\phi}$ can be seen
to be a solution of Eq.\ \eqref{eq:mathieu} provided that
\begin{equation}
  (m-\mu)^2 = \frac{E}{\epsilon_0} + \frac{\alpha^2}{4\epsilon_0^2a^2}.
  \label{eq:mu_1}
\end{equation}

Inserting now the spinor part into Eq.\ \eqref{eq:chi} we get
\begin{equation}
  \frac{c_{\uparrow}}{c_{\downarrow}} = \frac{\alpha}{2\epsilon_0a\mu} =
  \frac{2\epsilon_0a(\mu + 1)}{\alpha}
  \label{eq:chi-up-down}
\end{equation}
which can be solved for $\mu$ to yield
\begin{equation}
  \mu_{\pm} = -\frac{1}{2} \pm \frac{1}{2}\sqrt{1 + \frac{\alpha^2}{\epsilon_0^2a^2}}.
  \label{eq:mu_2}
\end{equation}
These exponents can be rewritten as $\mu_{\pm} = \pm\mu + n_{\pm}$ in terms of a
characteristic exponent $\mu$, which satisfies $|\mu|\leq1/2$, and a pair of integers
$n_{\pm}$, if the former and the integer $n_+$ are defined through the equality
\begin{equation}
  \mu_+ - \mu_- = \sqrt{1 + \frac{\alpha^2}{\epsilon_0^2a^2}} = 2n_+ + 1 + 2\mu.
\end{equation}

Equating the above expression to \eqref{eq:mu_1} and rearranging terms, the energy
spectrum is obtained
\begin{equation}
  \frac{E}{\epsilon_0} = \left(m + \frac{1}{2}\right)^2 \mp\left(m + \frac{1}{2}\right)
  \sqrt{1 + \frac{\alpha^2}{\epsilon_0^2a^2}} + \frac{1}{4}.
  \label{eq:rashba-spectrum}
\end{equation}
Notice that this expression can also be obtained from the first two terms of the expansion
given in Eq. \eqref{eq:spectrum} which are independent of $|q|$ and correspond to the
exact form of the spectrum in this limit $q=0$.

The constants $\epsilon_0$ and $\alpha^2/\epsilon_0^2a^2$ can be seen to correspond
respectively to $\hbar\omega_0/2$ and $Q_R$ as defined in Ref.\ [\onlinecite{fru-ric}].
Moreover, $m$ can be decomposed into $m = \lambda n$, with $\lambda =\pm 1$ and
$n\in\mathbb{N}_0$, and the constants $c_{\uparrow}$ and $c_{\downarrow}$ can be seen
to depend, through $\mu$, only on the parameters of the system.
In turn, the factor $\mp(m + 1/2)$ in Eq. \eqref{eq:rashba-spectrum} can be rewritten as
$s|m + 1/2|$, where $s=\pm 1$.
%

Finally, it is worth noting that the same steps can be taken to arrive at a similar solution
in the pure Dresselhaus case $\alpha = 0\neq\beta$.

\subsection{Case $|\alpha| = |\beta|$}
\label{sec:equal}
It has been shown in a previous work\cite{sch-egu-los} that, in each of these cases ($\alpha = \pm\beta$), 
a conserved quantity appears which is associated to an equilibrium orientation of the spin
with respect to the axis of the ring.
These quantities can be used to obtain analytical expressions for the spinors
$\chi_{\mu\pm}(\phi)$ in both cases.


%
In what follows, we derive these expressions and from their periodicity we deduce the
corresponding Mathieu functions of the eigenstates of $H_{\phi}$.
For brevity and clarity, we concentrate on the case $\alpha = \beta$, but there is
actually no restriction to applying the same procedure to the $\alpha = -\beta$ case as
well.

We begin by computing the commutator
\begin{equation}
  \left[\alpha S^- -i\beta S^+,\alpha S^+ + i\beta S^-\right] = (\alpha^2 - \beta^2),
\end{equation}
which vanishes when $|\alpha|=|\beta|$.
This shows that, in those cases, the operators $\alpha S^- -i\beta S^+$ and its hermitian
conjugate can be simultaneously diagonalised. In the basis of eigenstates of the Pauli
matrix $\sigma_z$, $\{\chi_{\uparrow},\chi_{\downarrow}\}$, their eigenvectors are given
by
\begin{equation}
  \xi_{\pm} = \frac{1}{\sqrt{2}}\left[\chi_{\uparrow} \pm e^{i\pi/4}\chi_{\downarrow}\right]
\end{equation}
and the eigenvalues of $S^- -iS^+$ by $\lambda_{\pm} = \pm e^{-i\pi/4}$.
We thus propose and insert into Eq. \eqref{eq:chi} a solution of the form
\begin{equation}
  \chi_{\pm}(\phi) = g(\phi)\xi_{\pm}
\end{equation}
with $g(\phi)$ is a complex-valued scalar function, and get
\begin{equation}
  g'\xi_{\pm} = \frac{i\alpha}{2\epsilon_0a}\left[e^{i\phi}(S^- -iS^+) + e^{-i\phi}(S^+
    + iS^-)\right]g\xi_{\pm} = \pm\frac{i\alpha}{\epsilon_0a}\cos(\phi-\phi_0)g\xi_{\pm}
\end{equation}
where $\phi_0 = \pi/4$. This equation can be readily integrated to yield
\begin{equation}
  \chi_{\pm}(\phi) =
  \exp\left(\pm\frac{i\alpha}{a\epsilon_0}\sin(\phi-\phi_0)\right)\xi_{\pm} =
  \exp\left(\mp\frac{i\alpha}{\sqrt{2}a^2\epsilon_0}(x - y)\right)\xi_{\pm}
  \label{eq:chi_pm_equal}
\end{equation}
where the last equality is obtained by noting that if $x,y\in\mathbb{R}$ are defined as $x
+ iy = ae^{i\phi}$, then $a\sin(\phi - \pi/4) = (y-x)/\sqrt{2}$. Written as in
Eq. \eqref{eq:chi_pm_equal}, the expressions of $\chi_{\pm}(\phi)$ can be seen to
correspond to those given Ref.\ [\onlinecite{sch-egu-los}].

%
The $2\pi$-periodicity of $\chi_{\mu}(\phi)$ show that the degenerate states in this are
of form given in \eqref{eq:eigenspinor-periodic},
\begin{equation}
  \eta_{\pm,m}(\phi;q) = f_m\left(\phi + \frac{\pi}{4}\right)
  \exp\left(\pm\frac{i\alpha}{a\epsilon_0}\sin\left(\phi-\frac{\pi}{4}\right)\right)
  \frac{1}{\sqrt{2}}\begin{pmatrix}
    1 \\ \pm e^{i\pi/4}
  \end{pmatrix}
\end{equation}
where $f_m(\phi + \pi/4;q)$ are defined in Eq. \eqref{eq:fm-periodic}.



%
The energy of each eigenstates depends on $q = (\alpha/{2\epsilon_0a})^2$ and the order
$m$ of the Mathieu function $f_m(\phi;q)$.
For the ground states $m=0$ in particular, the spectrum can be expanded\cite{wol} as a power series
in $q$ that, up to the fourth-order, results
\begin{equation}
  \frac{E}{\epsilon_0} = -2q -\frac{1}{2}q^2 + \frac{7}{128}q^4 + O(q^6).
\end{equation}
%



\section{General Rashba+Dresselhaus case}
\label{sec:general-case}

In this section we drop the term $H_{\Delta}$ by setting $\Delta=0$ and concentrate on
those cases that correspond to the presence of both RSOI and DSOI with coupling constants
of arbitrary strength.
We analyze quantitatively the behavior of the eigenspinors $\eta_{\mu\pm}(\phi;q)$ and
their energy spectrum by integrating Eq. \eqref{eq:chi} for different combinations of the
dimensionless parameters $\bar{\alpha}=\alpha/2a\epsilon_0$ and
$\bar{\beta}=\beta/2a\epsilon_0$.

\begin{figure}[htbp]
  \includegraphics[scale=0.8]{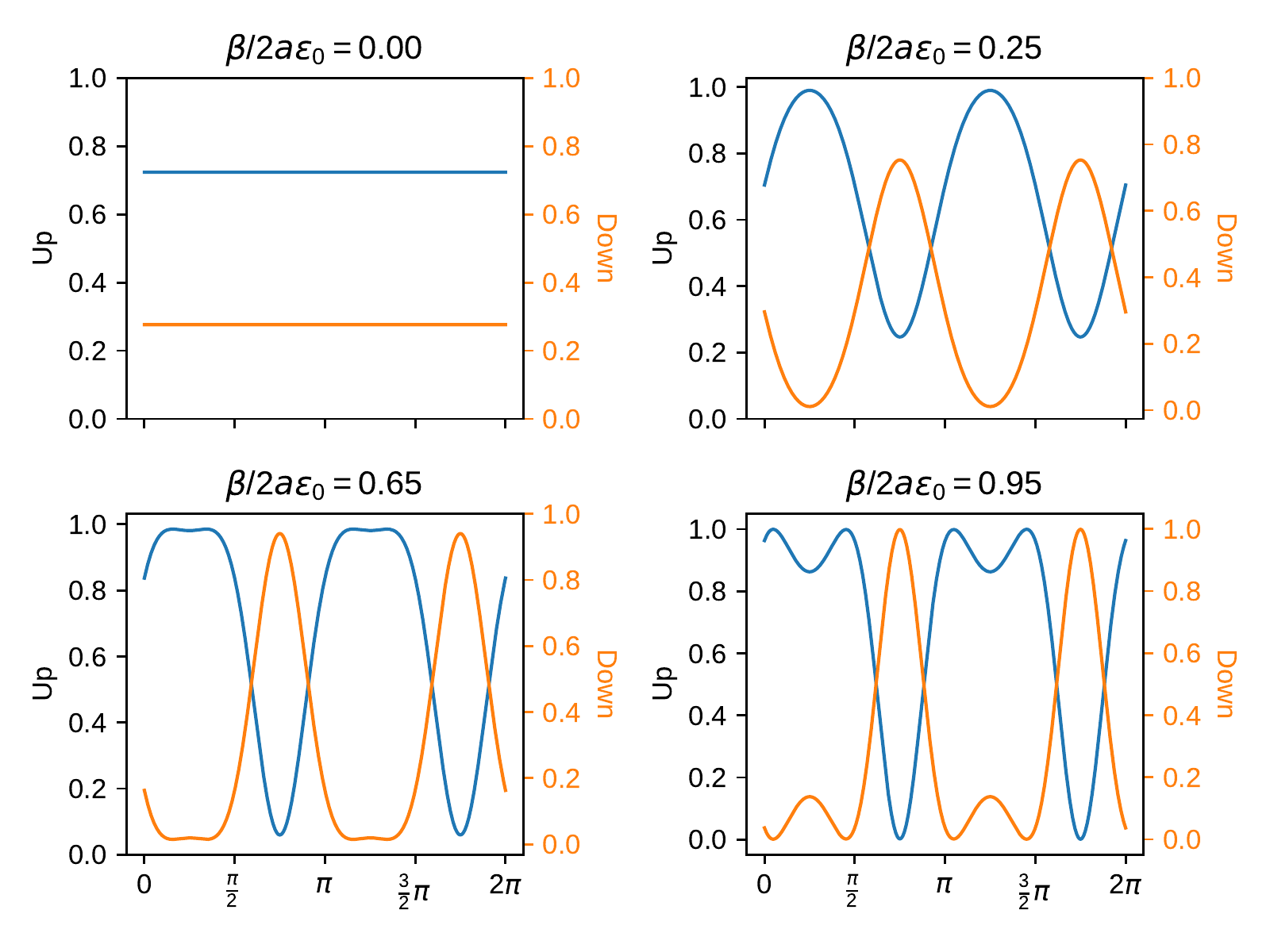}
  \caption{Behavior of the amplitude squared of the projections of the spinor
    $\chi_{\mu+}(\phi)$ onto the eigenstates of $\sigma_z$ for different values of
    $\beta/2a\epsilon_0$, while keeping $\alpha/2a\epsilon_0 = 1$. In all four cases, it
    can be seen that the spinor is normalized to unity for all $0\leq\phi\leq2\pi$. In the
    special case when $\beta=0$, the amplitude of each of the components becomes
    independent of $\phi$, as the exact solution in the pure Rashba case\cite{fru-ric}
    predicts (see Sec.\ \ref{sec:only-alpha}).}
  \label{fig:chi-sweep}
\end{figure}

\begin{figure}[htbp]
  \includegraphics[scale=0.8]{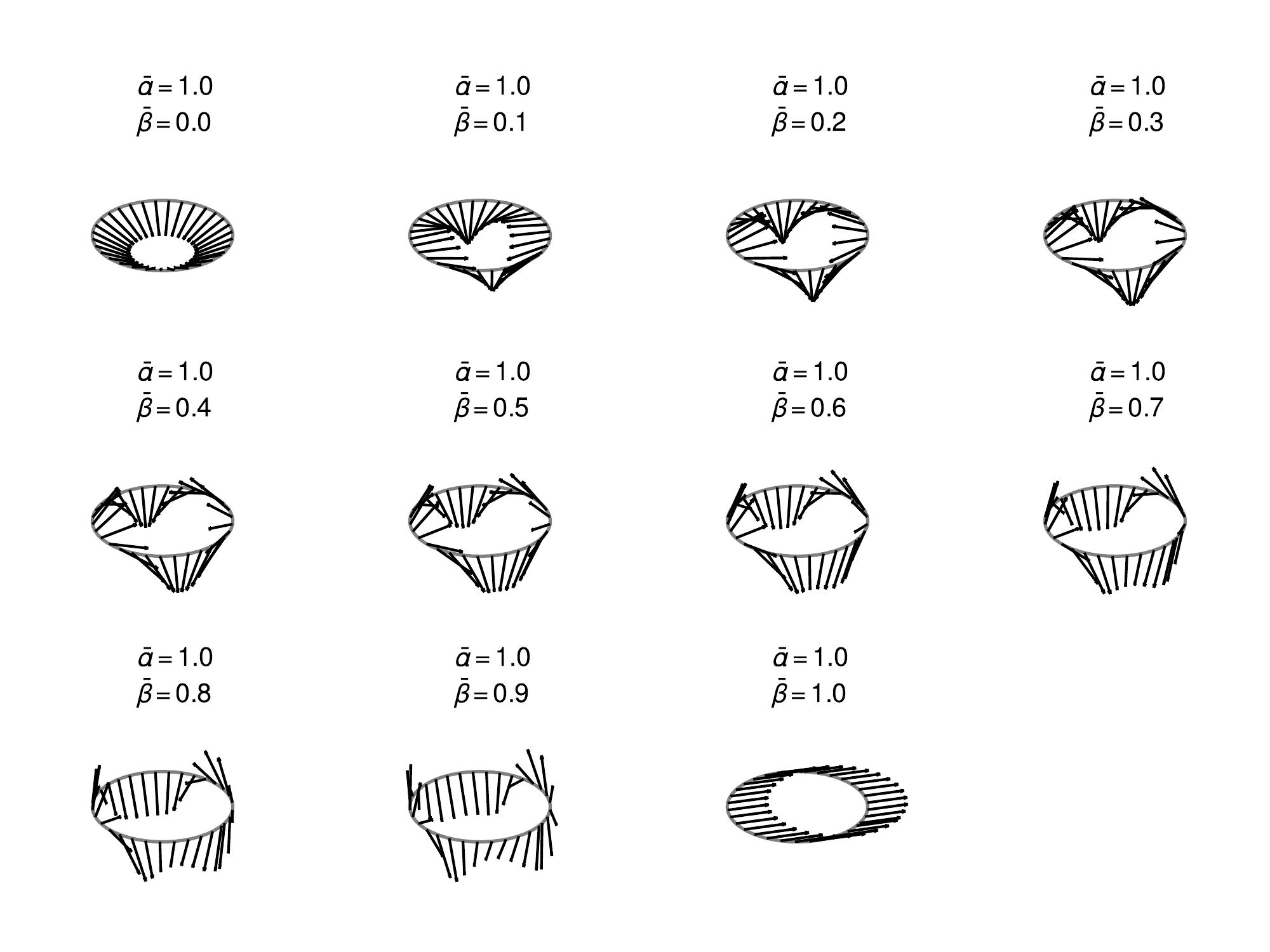}
  \caption{Bloch vector representation of the solution $\chi_{\mu+}(\phi)$ to
    Eq. \eqref{eq:chi} for different values of $\bar{\alpha}$ and $\bar{\beta}$. In the
    pure Rashba case $\bar{\beta}=0$, the orientation of the vector with respect to the
    axis of the ring is constant though its direction is not. It is in the case
    $|\bar{\alpha}| =|\bar{\beta}|$, that both its orientation and direction become
    independent of $\phi$. In these two cases the spin behaves as
    expected\cite{fru-ric,sch-egu-los} (see Secs. \ref{sec:only-alpha} and
    \ref{sec:equal}).}
  \label{fig:chi-arrows}
\end{figure}

In Figs. \ref{fig:chi-sweep} and \ref{fig:chi-arrows}, we set $\bar{\alpha}=1$ and compute
for different values of $\bar{\beta}$ the amplitude squared of the projections onto the
eigenstates of $\sigma_z$ and the graphical representation of the Bloch vector associated
with the solution $\chi_{\mu+}(\phi)$.


%
%

\begin{figure}[htbp]
  \includegraphics[scale=0.8]{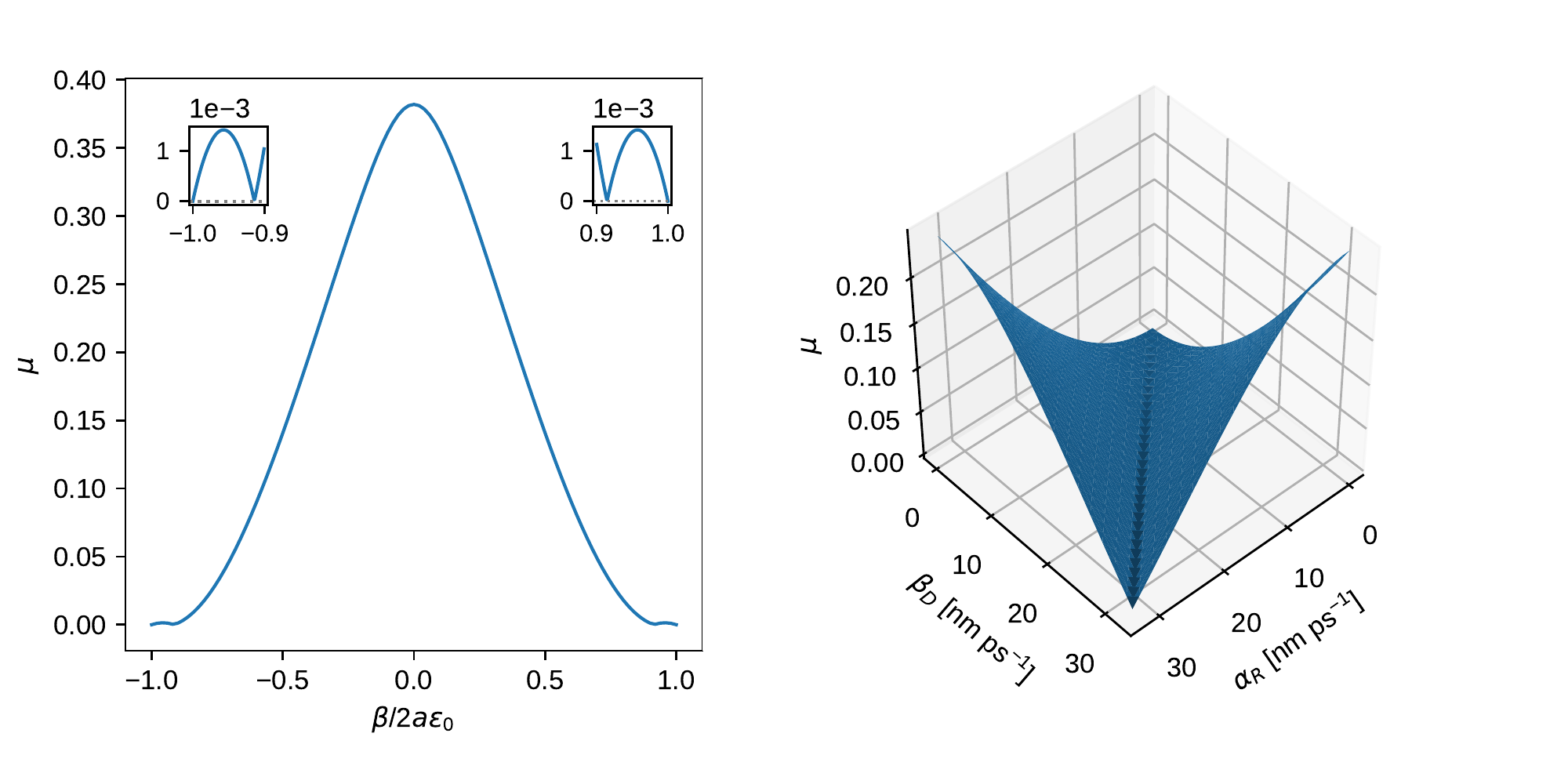}
  \caption{Left: dependence of $\mu$ on $\bar{\beta}$ while keeping
    $\bar{\alpha} = 1$. Notice that $\mu$ reaches the special value $\mu=0$ at the
    points $\bar{\alpha}=\pm\bar{\beta}$, as it is predicted by the analytical solution to
    the problem in these cases\cite{sch-egu-los} (see Sec. \ref{sec:equal}). 
    The sharp change in the behavior of $\mu$ for values of $\bar{\beta}$ in the range
    $0.9\leq|\bar{\beta}|\leq1$ seems to suggest the existence of another two zeros. This,
    however, can be ruled out by numerically estimating both minimums.
    %
    %
    A similar behavior is obtained if
    $\bar{\beta}$ is kept fixed and $\bar{\alpha}$ is allowed to vary.  Right: behavior of
    $\mu$ as a function of $\alpha_R$ and $\beta_D$ for realistic values of these
    parameters and for a ring of radius 200~\AA{} with a conduction-band effective mass of
    $0.1m_0$ where $m_0$ is the bare electron mass.}
  \label{fig:mu}
\end{figure}

In Fig.\ \ref{fig:mu}, we study the dependence of the Floquet exponent $\mu$ on
$\bar{\beta}$ on them first by setting one parameter to unity and letting the other vary
continously, and then by choosing a range of realisitic values for the parameters
$\alpha_R = \alpha/\hbar$ and $\beta_D = \beta/\hbar$.

\begin{figure}[htbp]
  \includegraphics[scale=0.8]{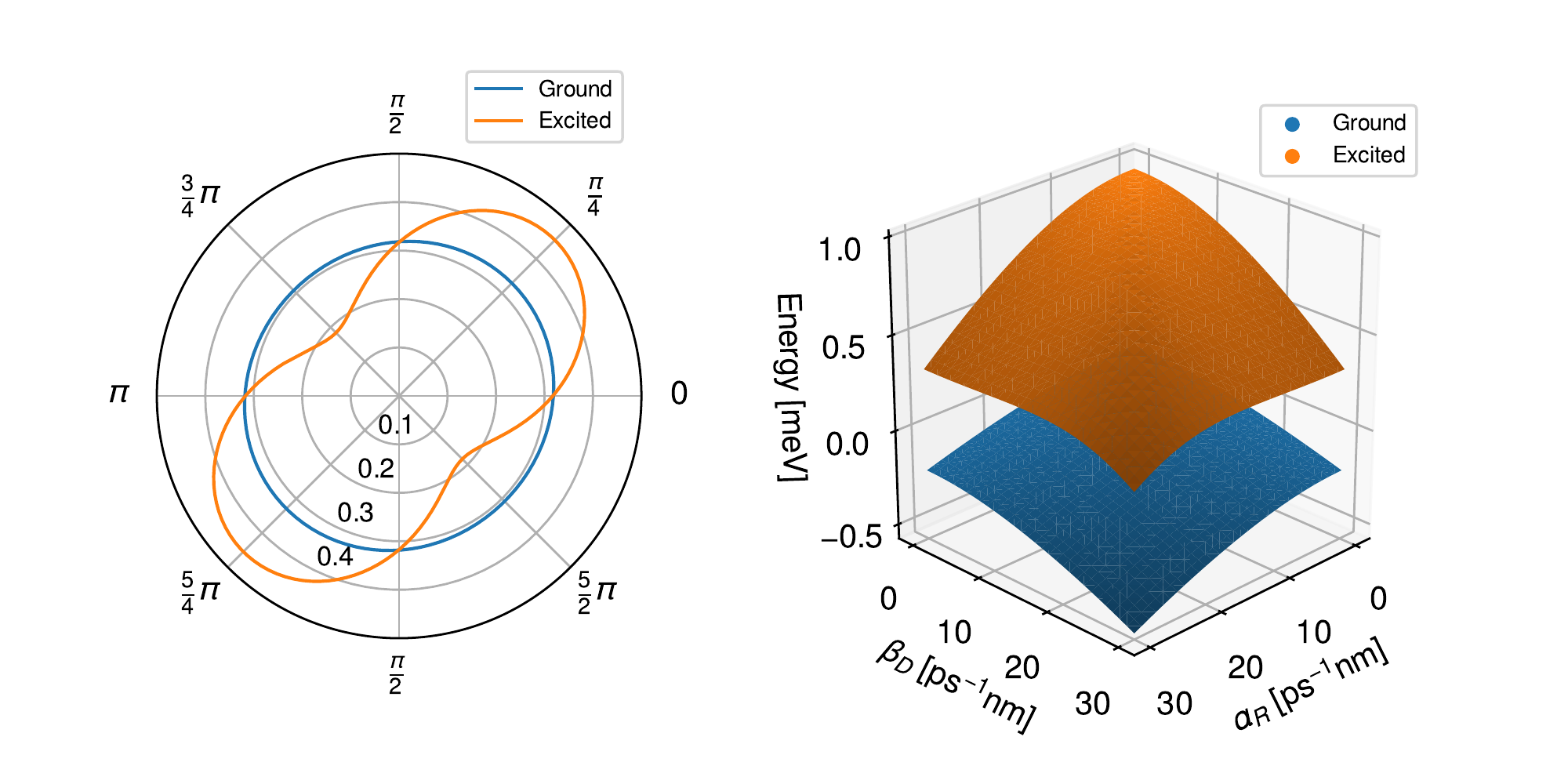}
  \caption{Left: density of the degenerate ground state $\eta_{\mu+,0}(\phi;q)$ and first
    excited state $\eta_{\mu+,1}(\phi;q)$ for a ring of radius 200~\AA{}, a conduction
    band effective mass of $0.1m_0$, with $m_0$ the bare electron mass, and the SO
    coupling constants $\alpha_R = 8~\mathrm{nm}\,\mathrm{ps}^{-1}$ and $\beta_D =
    14~\mathrm{nm}\,\mathrm{ps}^{-1}$ ($q\approx 0.033$).  Right: the spectrum for the
    ground and excited states as a function of $\alpha_R$ and $\beta_D$. Notice that, when
    $\alpha_R = \beta_D = 0$, the energies of the ground and the excited states tend to
    zero and $\epsilon_0\approx 0.95~\mathrm{meV}$, respectively. These results are
    consistent with the exact solution (see Sec. \ref{sec:free-electron}).}
  \label{fig:mathieu}
\end{figure}

Finally, in Fig. \ref{fig:mathieu} we analyze the dependence of the spectrum on the
parameters $\alpha_R$ and $\beta_D$ for values in the same range explored in
Fig. \ref{fig:mu}. We also compute the density of the ground and excited states,
$|\eta_{\mu+,0}(\phi;q)|^2$ and $|\eta_{\mu+,1}(\phi; q)|^2$, respectively.

\section{Conclusion}
\label{sec:conclusion}

We studied the energy eigenvalue problem of a charge in narrow,
quasi-one-dimensional semiconductor quantum rings in which the effects of spin-orbit
interaction are taken into account. 
We considered the usual\cite{sch-egu-los,fru-ric}
expressions for the Rashba and Dresselhaus SOI in this kind of geometry and included an
ad-hoc term that takes into account deviations from these two important interactions. 
We found a factorization of the problem that does not make use of approximations and can be
applied to derive expressions for both the eigenstates of the full Hamiltonian and the
energy spectrum. 
In Sec.\ \ref{sec:solution}, we showed that each eigenstate can be
written as a product of spinor function, which is a Floquet solution to
Eq.\ \eqref{eq:chi}, and a scalar Mathieu function whose Floquet multiplier is chosen
conjugate to that of the spinor factor. 
This relation is important because the real parts
of the Floquet exponents associated with this multiplier determine the orders of the
Mathieu functions and the spectrum of energies.
With recourse to Mathieu's equation theory, we found that the eigenstates so obtained are at
least doubly degenerate. 
This degeneracy is also present in the special cases discussed in
Secs.\ \ref{sec:free-electron}, \ref{sec:only-alpha} and \ref{sec:equal}, and is
accounted for by the time-reversal invariance of the Hamiltonian. 
Finally, we showed in the aforementioned sections that the obtained expressions of the eigenstates 
and the energy spectrum reduce to those already known in the literature when appropriate 
limits are taken.


\section{Acknowledgements}

We gratefully acknowledge financial support from  UBACyT of the Universidad de Buenos Aires, 
and from CONICET.

\appendix
\section{}
\label{sec:appendix}
Provided that the angular variable $\phi$ is identified with ``time'', the anti-hermitian
property of the operator $F(\phi)$ allows Eq. \eqref{eq:chi} to be interpreted as a
time-dependent Schr\"odinger equation
\begin{equation}
  i\chi' = \frac{i}{2\epsilon_0}F(\phi)\chi.
  \label{eq:chi-schroedinger}
\end{equation}
where the hermitian operator $iF(\phi)/2\epsilon_0$ can be thought of as its ``Hamiltonian''.
It is well-known from quantum mechanics\cite{sak-nap} that any solution $\chi(\phi)$ to this
equation can be expressed in terms of a constant spinor $\chi(0)$ and an unitary evolution
operator $U(\phi)$ that is also a solution to Eq. \eqref{eq:chi}, as
\begin{equation}
  \chi(\phi) = U(\phi)\chi(0).
  \label{eq:chi-sol}
\end{equation}
With recourse to Floquet theory, it can also be shown that, in the cases we are
considering, the $2\pi$-periodicity of $F(\phi)$ endows the evolution operator $U(\phi)$
with the pseudo-periodic property
\begin{equation}
  U(\phi + 2\pi) = U(\phi)U(2\pi),
\end{equation}

As $U(2\pi)$ is also unitary, its eigenvectors form an orthonormal set. Therefore, a pair
of orthonormal pseudo-periodic solutions to Eq. \eqref{eq:chi-schroedinger} can be readily
constructed by applying the evolution operator $U(\phi)$ to the (constant) eigenvectors of
$U(2\pi)$. More specifically, if $\chi_{\rho}(0)$ is an eigenvector of $U(2\pi)$ with
eigenvalue $\rho$, then there exists a solution $\chi_{\rho}(\phi)$ of the form
\eqref{eq:chi-sol} that satisfies
\begin{equation}
  \chi_{\rho}(\phi + 2\pi) = U(\phi + 2\pi)\chi_{\rho}(0) = U(\phi)U(2\pi)\chi_{\rho}(0) =
  \rho U(\phi)\chi_{\rho}(0) = \rho\chi_{\rho}(\phi).
\end{equation}
The unitarity of $U(2\pi)$ implies that all of its eigenvalues satisfy $|\rho| =
1$ and therefore are of the form $e^{i\gamma}$, with $\gamma\in\mathbb{R}$.
It can also be shown, by computing the determinant of $U(2\pi)$ through Jacobi's formula,
that its (two) eigenvalues $\rho_{\pm}$ are related by
\begin{equation}
  \rho_{+}\rho_{-} = \det U(2\pi) = \exp\left(\frac{1}{2\epsilon_0}\int_0^{2\pi}\Tr
  F(\phi')\,\mathrm{d}\phi'\right) = 1,
\end{equation}
since $\Tr S^{\pm} = 0$ and therefore $\Tr F(\phi) = 0$.
This relation, together with $|\rho_{\pm}|=1$, shows that these eigenvalues are of the
form $\rho_{\pm} = e^{2\pi i(\pm\mu + n_{\pm})}$, with $0\leq\mu\leq1/2$ and
$n_{\pm}\in\mathbb{Z}$. The quantities $\rho_{\pm}$ are the characteristic Floquet
multipliers of the system and $i(\pm\mu + n_{\pm})$ their Floquet exponents.

Finally, it stems from the unitarity of $U(2\pi)$ and the orthonormality of its
eigenvectors $\chi_{\rho}(0)$ that the solutions $\chi_{\rho}(\phi)$ are orthonormal for
all $\phi$, since
\begin{equation}
  \chi_{\rho}(\phi)^\dag\chi_{\rho'}(\phi) =
      [U(\phi)\chi_{\rho}(0)]^\dag[U(\phi)\chi_{\rho'}(0)] = \chi_{\rho}(0)^\dag
      U(\phi)^\dag U(\phi)\chi_{\rho'}(0) = \chi_{\rho}(0)^\dag\chi_{\rho'}(0) =
      \delta_{\rho\rho'}.
\end{equation}

\section{}
In what follows, we mention some of the properties of Mathieu functions of both integer
and non-integer order that relate directly to their orthonormality in the interval
$0\leq\phi\leq2\pi$.
For proofs and detailed derivations, we refer the reader to
Refs. \onlinecite{wol,mcl,ars}.

It can be shown that, when $q,\phi\in\mathbb{R}$, the even $\ce_m(\phi;q)$ ($m\geq0$) and odd
$\se_m(\phi;q)$ ($m>0$) Mathieu functions of integer order are real and satisfy
\begin{equation}
  \begin{aligned}
    \int_0^{2\pi} \ce_m(\phi;q)\ce_n(\phi;q)\,d\phi &= \pi\delta_{mn}, \\
    \int_0^{2\pi} \se_m(\phi;q)\se_n(\phi;q)\,d\phi &= \pi\delta_{mn}; \\
    \int_0^{2\pi} \ce_m(\phi;q)\se_n(\phi;q)\,d\phi &= 0; \\
  \end{aligned}
\end{equation}
where $\delta_{mn}$ is the Kronecker delta.

In turn, it can be shown that, if $\nu,\phi,q\in\mathbb{R}$, Mathieu functions
$\me_{\nu}(\phi;q)$ satisfy
\begin{align}
  \me_{\nu}(\phi;q)^\ast &= \me_{\nu}(-\phi;q), \label{eq:prop_b1}\\
  \me_{\nu}(\phi + \pi;q) &= e^{i\nu\pi}\me_{\nu}(\phi;q),\label{eq:prop_b2} \\
  \int_0^{\pi}\me_{\nu + 2n}(-\phi;q)&\me_{\nu + 2m}(\phi;q)\,d\phi =
  \pi\delta_{nm};\label{eq:prop_b3}
\end{align}
where $n,m\in\mathbb{Z}$.
These three properties can be applied to write any well-behaved
pseudo-periodic function $f_{\nu}(\phi + \pi) = e^{i\nu\pi}f_{\nu}(\phi)$ as a series in
terms of the $\me_{\nu}(\phi;q)$.\cite{wol}

As to how properties \eqref{eq:prop_b2} and \eqref{eq:prop_b3} change when the interval is
extended to $0\leq\phi\leq2\pi$, it can be readily seen that the former implies
$\me_{\nu}(\phi + 2\pi;q) = e^{2\pi i\nu}\me_{\nu}(\phi;q)$, whereas for the latter one
can show, using the Fourier series of the functions $\me_{\nu}(\phi;q)$, that it takes the
form
%
%
\begin{equation}
  \int_0^{2\pi}\me_{-(\nu + n)}(\phi;q)\me_{\nu + m}(\phi;q)\,d\phi = 2\pi\delta_{mn}.
\end{equation}
This relation, together with property \eqref{eq:prop_b1}, allows the product $\me_{-(\nu +
  m)}(\phi;q)\me_{\nu + m}(\phi;q)$ to be interpreted as a density.
  
\bibliography{references.bib}

\begin{thebibliography}{20}%
\makeatletter
\providecommand \@ifxundefined [1]{%
 \@ifx{#1\undefined}
}%
\providecommand \@ifnum [1]{%
 \ifnum #1\expandafter \@firstoftwo
 \else \expandafter \@secondoftwo
 \fi
}%
\providecommand \@ifx [1]{%
 \ifx #1\expandafter \@firstoftwo
 \else \expandafter \@secondoftwo
 \fi
}%
\providecommand \natexlab [1]{#1}%
\providecommand \enquote  [1]{``#1''}%
\providecommand \bibnamefont  [1]{#1}%
\providecommand \bibfnamefont [1]{#1}%
\providecommand \citenamefont [1]{#1}%
\providecommand \href@noop [0]{\@secondoftwo}%
\providecommand \href [0]{\begingroup \@sanitize@url \@href}%
\providecommand \@href[1]{\@@startlink{#1}\@@href}%
\providecommand \@@href[1]{\endgroup#1\@@endlink}%
\providecommand \@sanitize@url [0]{\catcode `\\12\catcode `\$12\catcode
  `\&12\catcode `\#12\catcode `\^12\catcode `\_12\catcode `\%12\relax}%
\providecommand \@@startlink[1]{}%
\providecommand \@@endlink[0]{}%
\providecommand \url  [0]{\begingroup\@sanitize@url \@url }%
\providecommand \@url [1]{\endgroup\@href {#1}{\urlprefix }}%
\providecommand \urlprefix  [0]{URL }%
\providecommand \Eprint [0]{\href }%
\providecommand \doibase [0]{http://dx.doi.org/}%
\providecommand \selectlanguage [0]{\@gobble}%
\providecommand \bibinfo  [0]{\@secondoftwo}%
\providecommand \bibfield  [0]{\@secondoftwo}%
\providecommand \translation [1]{[#1]}%
\providecommand \BibitemOpen [0]{}%
\providecommand \bibitemStop [0]{}%
\providecommand \bibitemNoStop [0]{.\EOS\space}%
\providecommand \EOS [0]{\spacefactor3000\relax}%
\providecommand \BibitemShut  [1]{\csname bibitem#1\endcsname}%
\let\auto@bib@innerbib\@empty
\bibitem [{\citenamefont {Fomin}(2014)}]{fom}%
  \BibitemOpen
  \bibfield  {author} {\bibinfo {author} {\bibfnamefont {V.~M.}\ \bibnamefont
  {Fomin}},\ }\href@noop {} {\emph {\bibinfo {title} {Physics of {Q}uantum
  {R}ings}}},\ NanoScience and Technology\ (\bibinfo  {publisher} {Springer},\
  \bibinfo {year} {2014})\BibitemShut {NoStop}%
\bibitem [{\citenamefont {Cygorek}\ \emph {et~al.}(2015)\citenamefont
  {Cygorek}, \citenamefont {Tamborenea},\ and\ \citenamefont
  {Axt}}]{cyg-tam-axt}%
  \BibitemOpen
  \bibfield  {author} {\bibinfo {author} {\bibfnamefont {M.}~\bibnamefont
  {Cygorek}}, \bibinfo {author} {\bibfnamefont {P.~I.}\ \bibnamefont
  {Tamborenea}}, \ and\ \bibinfo {author} {\bibfnamefont {V.~M.}\ \bibnamefont
  {Axt}},\ }\href {\doibase 10.1103/PhysRevB.92.115301} {\bibfield  {journal}
  {\bibinfo  {journal} {Phys. Rev. B}\ }\textbf {\bibinfo {volume} {92}},\
  \bibinfo {pages} {115301} (\bibinfo {year} {2015})}\BibitemShut {NoStop}%
\bibitem [{\citenamefont {Quinteiro}\ \emph {et~al.}(2011)\citenamefont
  {Quinteiro}, \citenamefont {Tamborenea},\ and\ \citenamefont
  {Berakdar}}]{qui-tam-ber}%
  \BibitemOpen
  \bibfield  {author} {\bibinfo {author} {\bibfnamefont {G.~F.}\ \bibnamefont
  {Quinteiro}}, \bibinfo {author} {\bibfnamefont {P.~I.}\ \bibnamefont
  {Tamborenea}}, \ and\ \bibinfo {author} {\bibfnamefont {J.}~\bibnamefont
  {Berakdar}},\ }\href {\doibase 10.1364/OE.19.026733} {\bibfield  {journal}
  {\bibinfo  {journal} {Opt. Express}\ }\textbf {\bibinfo {volume} {19}},\
  \bibinfo {pages} {26733} (\bibinfo {year} {2011})}\BibitemShut {NoStop}%
\bibitem [{\citenamefont {K\"onig}\ \emph {et~al.}(2006)\citenamefont
  {K\"onig}, \citenamefont {Tschetschetkin}, \citenamefont {Hankiewicz},
  \citenamefont {Sinova}, \citenamefont {Hock}, \citenamefont {Daumer},
  \citenamefont {Sch\"afer}, \citenamefont {Becker}, \citenamefont {Buhmann},\
  and\ \citenamefont {Molenkamp}}]{koe-tsc-han}%
  \BibitemOpen
  \bibfield  {author} {\bibinfo {author} {\bibfnamefont {M.}~\bibnamefont
  {K\"onig}}, \bibinfo {author} {\bibfnamefont {A.}~\bibnamefont
  {Tschetschetkin}}, \bibinfo {author} {\bibfnamefont {E.~M.}\ \bibnamefont
  {Hankiewicz}}, \bibinfo {author} {\bibfnamefont {J.}~\bibnamefont {Sinova}},
  \bibinfo {author} {\bibfnamefont {V.}~\bibnamefont {Hock}}, \bibinfo {author}
  {\bibfnamefont {V.}~\bibnamefont {Daumer}}, \bibinfo {author} {\bibfnamefont
  {M.}~\bibnamefont {Sch\"afer}}, \bibinfo {author} {\bibfnamefont {C.~R.}\
  \bibnamefont {Becker}}, \bibinfo {author} {\bibfnamefont {H.}~\bibnamefont
  {Buhmann}}, \ and\ \bibinfo {author} {\bibfnamefont {L.~W.}\ \bibnamefont
  {Molenkamp}},\ }\href {\doibase 10.1103/PhysRevLett.96.076804} {\bibfield
  {journal} {\bibinfo  {journal} {Phys. Rev. Lett.}\ }\textbf {\bibinfo
  {volume} {96}},\ \bibinfo {pages} {076804} (\bibinfo {year}
  {2006})}\BibitemShut {NoStop}%
\bibitem [{\citenamefont {Bergsten}\ \emph {et~al.}(2006)\citenamefont
  {Bergsten}, \citenamefont {Kobayashi}, \citenamefont {Sekine},\ and\
  \citenamefont {Nitta}}]{ber-kob-sek}%
  \BibitemOpen
  \bibfield  {author} {\bibinfo {author} {\bibfnamefont {T.}~\bibnamefont
  {Bergsten}}, \bibinfo {author} {\bibfnamefont {T.}~\bibnamefont {Kobayashi}},
  \bibinfo {author} {\bibfnamefont {Y.}~\bibnamefont {Sekine}}, \ and\ \bibinfo
  {author} {\bibfnamefont {J.}~\bibnamefont {Nitta}},\ }\href {\doibase
  10.1103/PhysRevLett.97.196803} {\bibfield  {journal} {\bibinfo  {journal}
  {Phys. Rev. Lett.}\ }\textbf {\bibinfo {volume} {97}},\ \bibinfo {pages}
  {196803} (\bibinfo {year} {2006})}\BibitemShut {NoStop}%
\bibitem [{\citenamefont {Nagasawa}\ \emph {et~al.}(2013)\citenamefont
  {Nagasawa}, \citenamefont {Frustaglia}, \citenamefont {Saarikoski},
  \citenamefont {Richter},\ and\ \citenamefont {Nitta}}]{nag-fum-fru}%
  \BibitemOpen
  \bibfield  {author} {\bibinfo {author} {\bibfnamefont {F.}~\bibnamefont
  {Nagasawa}}, \bibinfo {author} {\bibfnamefont {D.}~\bibnamefont
  {Frustaglia}}, \bibinfo {author} {\bibfnamefont {H.}~\bibnamefont
  {Saarikoski}}, \bibinfo {author} {\bibfnamefont {K.}~\bibnamefont {Richter}},
  \ and\ \bibinfo {author} {\bibfnamefont {J.}~\bibnamefont {Nitta}},\ }\href
  {http://dx.doi.org/10.1038/ncomms3526} {\bibfield  {journal} {\bibinfo
  {journal} {Nature Communications}\ }\textbf {\bibinfo {volume} {4}},\
  \bibinfo {pages} {2526} (\bibinfo {year} {2013})}\BibitemShut {NoStop}%
\bibitem [{\citenamefont {Schliemann}\ \emph {et~al.}(2003)\citenamefont
  {Schliemann}, \citenamefont {Egues},\ and\ \citenamefont
  {Loss}}]{sch-egu-los}%
  \BibitemOpen
  \bibfield  {author} {\bibinfo {author} {\bibfnamefont {J.}~\bibnamefont
  {Schliemann}}, \bibinfo {author} {\bibfnamefont {J.~C.}\ \bibnamefont
  {Egues}}, \ and\ \bibinfo {author} {\bibfnamefont {D.}~\bibnamefont {Loss}},\
  }\href {\doibase 10.1103/PhysRevLett.90.146801} {\bibfield  {journal}
  {\bibinfo  {journal} {Phys. Rev. Lett.}\ }\textbf {\bibinfo {volume} {90}},\
  \bibinfo {pages} {146801} (\bibinfo {year} {2003})}\BibitemShut {NoStop}%
\bibitem [{\citenamefont {Romano}\ \emph {et~al.}(2005)\citenamefont {Romano},
  \citenamefont {Ulloa},\ and\ \citenamefont {Tamborenea}}]{rom-ull-tam}%
  \BibitemOpen
  \bibfield  {author} {\bibinfo {author} {\bibfnamefont {C.~L.}\ \bibnamefont
  {Romano}}, \bibinfo {author} {\bibfnamefont {S.~E.}\ \bibnamefont {Ulloa}}, \
  and\ \bibinfo {author} {\bibfnamefont {P.~I.}\ \bibnamefont {Tamborenea}},\
  }\href {\doibase 10.1103/PhysRevB.71.035336} {\bibfield  {journal} {\bibinfo
  {journal} {Phys. Rev. B}\ }\textbf {\bibinfo {volume} {71}},\ \bibinfo
  {pages} {035336} (\bibinfo {year} {2005})}\BibitemShut {NoStop}%
\bibitem [{\citenamefont {Maiti}(2011)}]{mai}%
  \BibitemOpen
  \bibfield  {author} {\bibinfo {author} {\bibfnamefont {S.~K.}\ \bibnamefont
  {Maiti}},\ }\href@noop {} {\bibfield  {journal} {\bibinfo  {journal} {Journal
  of Applied Physics}\ }\textbf {\bibinfo {volume} {110}},\ \bibinfo {pages}
  {064306} (\bibinfo {year} {2011})}\BibitemShut {NoStop}%
\bibitem [{\citenamefont {Shakouri}\ \emph {et~al.}(2012)\citenamefont
  {Shakouri}, \citenamefont {Szafran}, \citenamefont {Esmaeilzadeh},\ and\
  \citenamefont {Peeters}}]{sha-sza-esm}%
  \BibitemOpen
  \bibfield  {author} {\bibinfo {author} {\bibfnamefont {K.}~\bibnamefont
  {Shakouri}}, \bibinfo {author} {\bibfnamefont {B.}~\bibnamefont {Szafran}},
  \bibinfo {author} {\bibfnamefont {M.}~\bibnamefont {Esmaeilzadeh}}, \ and\
  \bibinfo {author} {\bibfnamefont {F.~M.}\ \bibnamefont {Peeters}},\ }\href
  {\doibase 10.1103/PhysRevB.85.165314} {\bibfield  {journal} {\bibinfo
  {journal} {Phys. Rev. B}\ }\textbf {\bibinfo {volume} {85}},\ \bibinfo
  {pages} {165314} (\bibinfo {year} {2012})}\BibitemShut {NoStop}%
\bibitem [{\citenamefont {Zamani}\ \emph {et~al.}(2017)\citenamefont {Zamani},
  \citenamefont {Azargoshasb},\ and\ \citenamefont {Niknam}}]{zam-aza-nik}%
  \BibitemOpen
  \bibfield  {author} {\bibinfo {author} {\bibfnamefont {A.}~\bibnamefont
  {Zamani}}, \bibinfo {author} {\bibfnamefont {T.}~\bibnamefont {Azargoshasb}},
  \ and\ \bibinfo {author} {\bibfnamefont {E.}~\bibnamefont {Niknam}},\ }\href
  {\doibase https://doi.org/10.1016/j.physb.2017.08.031} {\bibfield  {journal}
  {\bibinfo  {journal} {Physica B: Condensed Matter}\ }\textbf {\bibinfo
  {volume} {523}},\ \bibinfo {pages} {85 } (\bibinfo {year}
  {2017})}\BibitemShut {NoStop}%
\bibitem [{\citenamefont {Pourmand}\ and\ \citenamefont
  {Rezaei}(2018)}]{pou-rez}%
  \BibitemOpen
  \bibfield  {author} {\bibinfo {author} {\bibfnamefont {S.~E.}\ \bibnamefont
  {Pourmand}}\ and\ \bibinfo {author} {\bibfnamefont {G.}~\bibnamefont
  {Rezaei}},\ }\href {\doibase https://doi.org/10.1016/j.physb.2018.04.046}
  {\bibfield  {journal} {\bibinfo  {journal} {Physica B: Condensed Matter}\
  }\textbf {\bibinfo {volume} {543}},\ \bibinfo {pages} {27 } (\bibinfo {year}
  {2018})}\BibitemShut {NoStop}%
\bibitem [{\citenamefont {Frustaglia}\ and\ \citenamefont
  {Richter}(2004)}]{fru-ric}%
  \BibitemOpen
  \bibfield  {author} {\bibinfo {author} {\bibfnamefont {D.}~\bibnamefont
  {Frustaglia}}\ and\ \bibinfo {author} {\bibfnamefont {K.}~\bibnamefont
  {Richter}},\ }\href {\doibase 10.1103/PhysRevB.69.235310} {\bibfield
  {journal} {\bibinfo  {journal} {Phys. Rev. B}\ }\textbf {\bibinfo {volume}
  {69}},\ \bibinfo {pages} {235310} (\bibinfo {year} {2004})}\BibitemShut
  {NoStop}%
\bibitem [{\citenamefont {Winkler}(2003)}]{win}%
  \BibitemOpen
  \bibfield  {author} {\bibinfo {author} {\bibfnamefont {R.}~\bibnamefont
  {Winkler}},\ }\href@noop {} {\emph {\bibinfo {title} {Spin-Orbit Coupling in
  Two-Dimensional Electron and Hole System}}},\ Vol.\ \bibinfo {volume} {191}\
  (\bibinfo  {publisher} {Springer Verlag},\ \bibinfo {address} {Berlin
  Heidelberg},\ \bibinfo {year} {2003})\BibitemShut {NoStop}%
\bibitem [{\citenamefont {Meijer}\ \emph {et~al.}(2002)\citenamefont {Meijer},
  \citenamefont {Morpurgo},\ and\ \citenamefont {Klapwijk}}]{mei-mor-klap}%
  \BibitemOpen
  \bibfield  {author} {\bibinfo {author} {\bibfnamefont {F.~E.}\ \bibnamefont
  {Meijer}}, \bibinfo {author} {\bibfnamefont {A.~F.}\ \bibnamefont
  {Morpurgo}}, \ and\ \bibinfo {author} {\bibfnamefont {T.~M.}\ \bibnamefont
  {Klapwijk}},\ }\href {\doibase 10.1103/PhysRevB.66.033107} {\bibfield
  {journal} {\bibinfo  {journal} {Phys. Rev. B}\ }\textbf {\bibinfo {volume}
  {66}},\ \bibinfo {pages} {033107} (\bibinfo {year} {2002})}\BibitemShut
  {NoStop}%
\bibitem [{\citenamefont {McLachlan}(1951)}]{mcl}%
  \BibitemOpen
  \bibfield  {author} {\bibinfo {author} {\bibfnamefont {N.}~\bibnamefont
  {McLachlan}},\ }\href
  {https://archive.org/stream/in.ernet.dli.2015.215548/2015.215548.Theory-And#page/n5}
  {\emph {\bibinfo {title} {Theory and application of {Mathieu} functions}}},\
  \bibinfo {edition} {1st}\ ed.\ (\bibinfo  {publisher} {Clarendon},\ \bibinfo
  {year} {1951})\BibitemShut {NoStop}%
\bibitem [{\citenamefont {Wolf}(2010)}]{wol}%
  \BibitemOpen
  \bibfield  {author} {\bibinfo {author} {\bibfnamefont {G.}~\bibnamefont
  {Wolf}},\ }in\ \href@noop {} {\emph {\bibinfo {booktitle} {{NIST} Handbook of
  Mathematical Functions}}},\ \bibinfo {editor} {edited by\ \bibinfo {editor}
  {\bibfnamefont {F.}~\bibnamefont {Olver}}, \bibinfo {editor} {\bibfnamefont
  {D.}~\bibnamefont {Lozier}}, \bibinfo {editor} {\bibfnamefont
  {R.}~\bibnamefont {Boisvert}}, \ and\ \bibinfo {editor} {\bibfnamefont
  {C.}~\bibnamefont {Clark}}}\ (\bibinfo  {publisher} {Cambridge University
  Press},\ \bibinfo {year} {2010})\BibitemShut {NoStop}%
\bibitem [{\citenamefont {Arscott}(1964)}]{ars}%
  \BibitemOpen
  \bibfield  {author} {\bibinfo {author} {\bibfnamefont {F.~M.}\ \bibnamefont
  {Arscott}},\ }\href@noop {} {\emph {\bibinfo {title} {Periodic Differential
  Equations. An Introduction to Mathieu, Lamé, and Allied Functions.}}},\
  \bibinfo {series} {International Series of Monographs in Pure and Applied
  Mathematics}, Vol.~\bibinfo {volume} {66}\ (\bibinfo  {publisher} {Pergamon
  Press, The Macmillan Co., New York},\ \bibinfo {year} {1964})\BibitemShut
  {NoStop}%
\bibitem [{\citenamefont {Tamir}\ and\ \citenamefont {Wang}(1965)}]{tam-wan}%
  \BibitemOpen
  \bibfield  {author} {\bibinfo {author} {\bibfnamefont {T.}~\bibnamefont
  {Tamir}}\ and\ \bibinfo {author} {\bibfnamefont {H.}~\bibnamefont {Wang}},\
  }\href {\doibase 10.6028/jres.069B.011} {\bibfield  {journal} {\bibinfo
  {journal} {J. Res. Natl. Inst. Stan.}\ }\textbf {\bibinfo {volume} {69B}},\
  \bibinfo {pages} {101} (\bibinfo {year} {1965})}\BibitemShut {NoStop}%
\bibitem [{\citenamefont {Sakurai}\ and\ \citenamefont
  {Napolitano}(2011)}]{sak-nap}%
  \BibitemOpen
  \bibfield  {author} {\bibinfo {author} {\bibfnamefont {J.~J.}\ \bibnamefont
  {Sakurai}}\ and\ \bibinfo {author} {\bibfnamefont {J.}~\bibnamefont
  {Napolitano}},\ }\href
  {https://www.pearson.com/us/higher-education/program/Sakurai-Modern-Quantum-Mechanics-2nd-Edition/PGM160720.html?tab=overview}
  {\emph {\bibinfo {title} {Modern Quantum Mechanics}}},\ \bibinfo {edition}
  {2nd}\ ed.\ (\bibinfo  {publisher} {Pearson Education},\ \bibinfo {year}
  {2011})\BibitemShut {NoStop}%
\end{thebibliography}%
\end{document}